\documentclass[twocolumn,description,aps]{revtex4-1}
\usepackage{graphicx,epstopdf,amsmath,amssymb,multirow,CJK,color}
\usepackage[german,english]{babel}

\begin{document}

\title{Quantum spin Hall effect in monolayer and bilayer TaIrTe$_{4}$}

\author{Peng-Jie Guo}
\author{Xiao-Qin Lu}
\author{Wei Ji}
\author{Kai Liu}\email{kliu@ruc.edu.cn}
\author{Zhong-Yi Lu}\email{zlu@ruc.edu.cn}

\affiliation{Department of Physics and Beijing Key Laboratory of Opto-electronic Functional Materials $\&$ Micro-nano Devices, Renmin University of China, Beijing 100872, China}
\date{\today}

\begin{abstract}
Generally, stacking two quantum spin Hall insulators gives rise to a trivial insulator. Here, based on first-principles electronic structure calculations, we confirm that monolayer TaIrTe$_{4}$ is a quantum spin Hall insulator and remarkably find that bilayer TaIrTe$_{4}$ is still a quantum spin Hall insulator. Theoretical analysis indicates that the covalent-like interlayer interaction in combination with the small bandgap at time-reversal invariant $\Gamma$ point results in new band inversion in bilayer TaIrTe$_{4}$, namely, the emergence of quantum spin Hall phase. Meanwhile, a topological phase transition can be observed by increasing the interlayer distance in bilayer TaIrTe$_{4}$. Considering that bulk TaIrTe$_{4}$ is a type-II Weyl semimetal, layered TaIrTe$_{4}$ thus provides an ideal platform to realize different topological phases at different dimensions.
\end{abstract}

\date{\today} \maketitle


$Introduction$: Topological properties of real materials have attracted much attention both experimentally and theoretically in recent years. Quantum spin Hall (QSH) insulator namely two-dimensional topological insulator has a bandgap in the bulk and the topologically protected metallic states at the edge~\cite{Hasan,Qi}. Theoretically, a QHS insulator has odd band inversions but there will be even band inversions after stacking of two QSH insulators, thus stacking of two QSH insulators together is usually considered to give rise to a trivial insulator. Nevertheless, in real materials, non-negligible even strong interlayer interaction may exist for two stacked QSH insulators, which may lead to broadening of valence and conduction bands. As a result, one may ask: is it possible that stacking two quantum spin Hall insulators still gives rise to a QSH insulator in real materials? If so, this will provide a new platform for studying the QSH effect with more adjustable freedom.

Among various layered topological materials, tellurides have been intensively studied. Layered WTe$_{2}$ is a type-II Weyl semimetal in bulk form~\cite{Soluyanov-2015} while its monolayer is a QSH insulator defined on a curved Fermi level~\cite{Qian2014}.
Bulk TaIrTe$_{4}$, sharing the same nonsymmorphic space group symmetry as WTe$_{2}$, was firstly proposed to be a type-II Weyl semimetal with the minimum four Weyl points under time-reversal symmetry constraint~\cite{Brink-PRB, Khim-PRB, Hasan-nc, Haubold-PRB}. A later joint theory and experiment study suggested that bulk TaIrTe$_4$ has twelve Weyl points and a pair of node lines protected by a mirror symmetry~\cite{Zhou-PRB}.
Meanwhile, a very recent study indicated that monolayer TaIrTe$_4$ is a QSH insulator~\cite{Dong-prb}. Considering that tellurides show abundant layer-dependent electronic structures \cite{Liu-PRM, Qiao-sb}, it is interesting to investigate the topological properties of TaIrTe$_{4}$ at the two-dimensional (2D) limit.

In this paper, based on first-principles electronic structure calculations, we confirm that monolayer TaIrTe$_{4}$ is a quantum spin Hall insulator. More interestingly, in contrast to general perception that stacking two quantum spin Hall insulator would give rise to a trivial insulator, we exceptionally find that bilayer TaIrTe$_{4}$ is still a quantum spin Hall insulator. We have further explored the underlying physical mechanism theoretically.


$Method$: The electronic structures of ultrathin film TaIrTe$_{4}$ were studied with the projector augmented wave (PAW) method~\cite{B-PRB, Kresse-PRB} as implemented in the VASP package~\cite{Kresse-1993, Kresse-1996, Kresse-J}. The Perdew-Burke-Ernzerhof (PBE) type exchange-correlation functional at the generalized gradient approximation (GGA) level was adopted to describe the interaction between ionic core and valence electrons~\cite{Perdew-PRL}. The kinetic energy cutoff of the plane-wave basis was set to 400 eV. The Gaussian smearing method with a width of 0.05 eV was utilized for the Fermi surface broadening. For the structural relaxation of bulk TaIrTe$_{4}$, the optB86b-vdW functional~\cite{vdw-2011} was adopted to account for the interlayer interactions and an 18$\times$6$\times$6 $k$-point mesh was used for the Brillouin zone (BZ) sampling. Both cell parameters and internal atomic positions were fully relaxed until the forces on all atoms were smaller than 0.01 eV/{\AA}. The calculated lattice parameters of bulk TaIrTe$_{4}$ are in good agreement with the experimental values~\cite{Mar-1992}. For the monolayer and bilayer calculations, the in-plane lattice constants were fixed to the relaxed bulk values, while the internal atomic positions were fully relaxed with an 18$\times$6$\times$1 $k$-point mesh for the surface BZ sampling.
The topological invariants and the edge states of monolayer and bilayer TaIrTe$_{4}$ were studied at the equilibrium structures by using the WANNIERTOOLS package~\cite{Wu-wannier}.

\begin{figure*}[!t]
\centering
\includegraphics[width=0.8\textwidth]{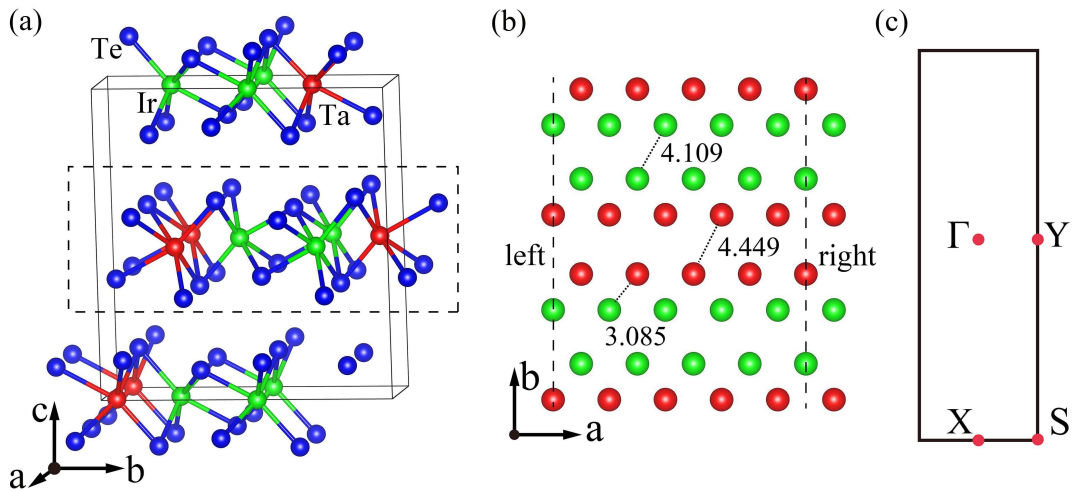}
\caption{(Color online) (a) Crystal structure of bulk TaIrTe$_{4}$. The red, green, and blue balls denote Ta, Ir, and Te atoms, respectively. The dashed rectangle highlights the layered structure. (b) Top view of monolayer TaIrTe$_{4}$, where Te atoms are hidden for clarity. Several typical interatomic distances are labeled. (c) Surface Brillouin zone (BZ) of monolayer TaIrTe$_{4}$. The red dots represent the high-symmetry $k$ points.}
\label{Fig1}
\end{figure*}


$Results$: Similar to type-II Weyl semimetal WTe$_{2}$, bulk TaIrTe$_{4}$ has a noncentrosymmetric crystal structure, adopting layered 1$T'$ structure and $AB$ stacking, as shown in Fig. \ref{Fig1}(a). Due to the structural distortion along $b$ axis, Ta and Ir atoms form zigzag chains along $a$ axis [Fig. \ref{Fig1}(b)].
The space group symmetry of bulk TaIrTe$_{4}$ is Pmn2$_{1}$, and the corresponding four point-group symmetric operations are identity E, mirror reflection M$_{x}$, screw operation $\{$C$_{2z}$$|$(1/2, 0, 1/2)$\}$, and glide reflection $\{$M$_{y}$$|$(1/2, 0, 1/2)$\}$, respectively.

\begin{figure}[!b]
\includegraphics[width=0.68\columnwidth]{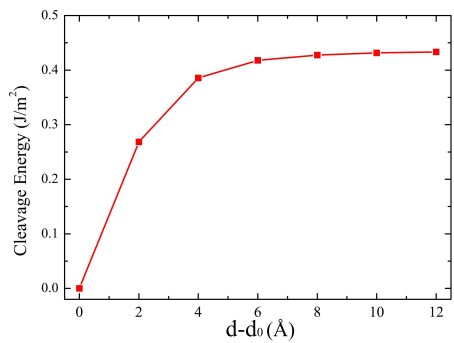} 
\caption{(Color online) Evolution of total energy of bulk TaIrTe$_{4}$ with interlayer distance $d$. Here $d_0$ denotes the equilibrium interlayer distance.}
\label{Fig2}
\end{figure}

Due to its layered structure, one would like to know whether or not TaIrTe$_{4}$ can be exfoliated as many other 2D materials. We calculated the energy of TaIrTe$_{4}$ as a function of the interlayer distance. As shown in Fig. \ref{Fig2}, the cleavage energy of TaIrTe$_{4}$ is 0.43 J/m$^{2}$, which is slightly larger than that of graphite (0.36 J/m$^{2}$)~\cite{Zacharia-PRB}. This indicates that atomically thin film of TaIrTe$_{4}$ is likely obtained by mechanical exfoliation as graphene.

\begin{figure}[!b]
\includegraphics[width=0.80\columnwidth]{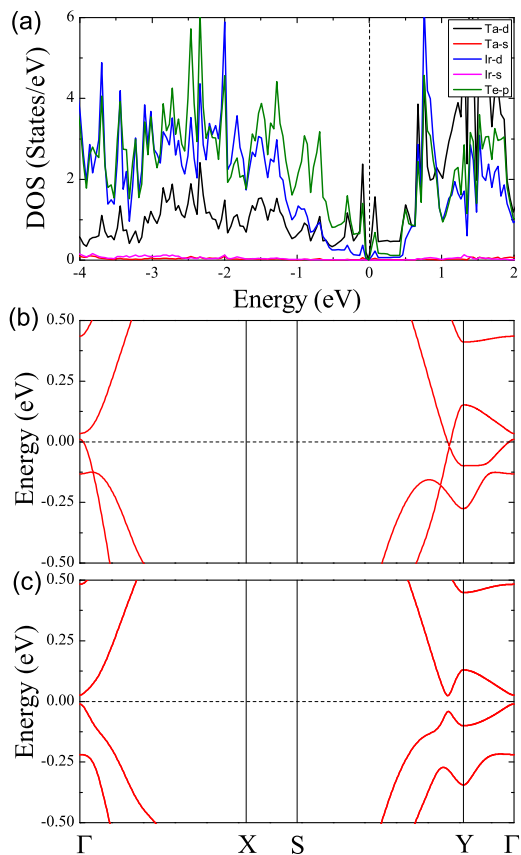}
\caption{(Color online) Electronic structure of monolayer TaIrTe$_{4}$: (a) density of states; band structures calculated (b) without and (c) with spin-orbit coupling.}
\label{Fig3}
\end{figure}

To study the possible topological property of bilayer TaIrTe4, we need first clarify the electronic structure of monolayerTaIrTe4. In comparison with bulk TaIrTe$_{4}$, the space group of monolayer TaIrTe$_{4}$ changes to P12$_{1}$/m1 with additional inversion symmetry, but losing nonsymmorphic symmetry operation of the fractional translation along $c$ axis. The corresponding surface Brillouin zone (SBZ) and the high-symmetry $k$ points are shown in Fig. \ref{Fig1}(c). From the calculated density of states (DOS) [Fig. \ref{Fig3}(a)], we can see that the states around the Fermi level are mainly contributed by Te $p$ orbitals and Ta $d$ orbitals, while Ir $d$ orbitals have relatively small contributions. The common peaks among these states indicate strong $p$-$d$ hybridization in monolayer TaIrTe$_{4}$.

The band structure of monolayer TaIrTe$_{4}$ along the high-symmetry paths of SBZ calculated without spin-orbit coupling (SOC) and with SOC are shown in Figs. \ref{Fig3}(b) and \ref{Fig3}(c), respectively. Without SOC, monolayer TaIrTe$_{4}$ is a semimetal, which has a band crossing along the S-Y path and a band inversion around the Y point [Fig. \ref{Fig3}(b)]. Moreover, the bands around the Fermi level have small dispersion along the $\Gamma$-Y path, which is vertical to the atomic chain direction ($a$ axis) in real space [Fig. \ref{Fig1}(b)]. We have also examined the band structure by using the hybrid functional with the HSE06 version~\cite{Krukau-2006}, which gives similar results. When including SOC, notable changes take place in the band structure. Monolayer TaIrTe$_{4}$ transforms to an insulator with a band gap of 32 meV.
  Meanwhile the band inversion around the time-reversal-invariant Y point may result in nontrivial topological property, we thus calculated the topological invariant Z$_{2}$ by using the Wilson loops method~\cite{Yu-PRB} and obtained the topological invariant Z$_{2}$ as 1. This demonstrates that monolayer TaIrTe$_4$ is a quantum spin Hall insulator~\cite{Dong-prb}, similar to monolayer WTe$_{2}$~\cite{Qian2014}.

\begin{figure}[!t]
\includegraphics[width=0.7\columnwidth]{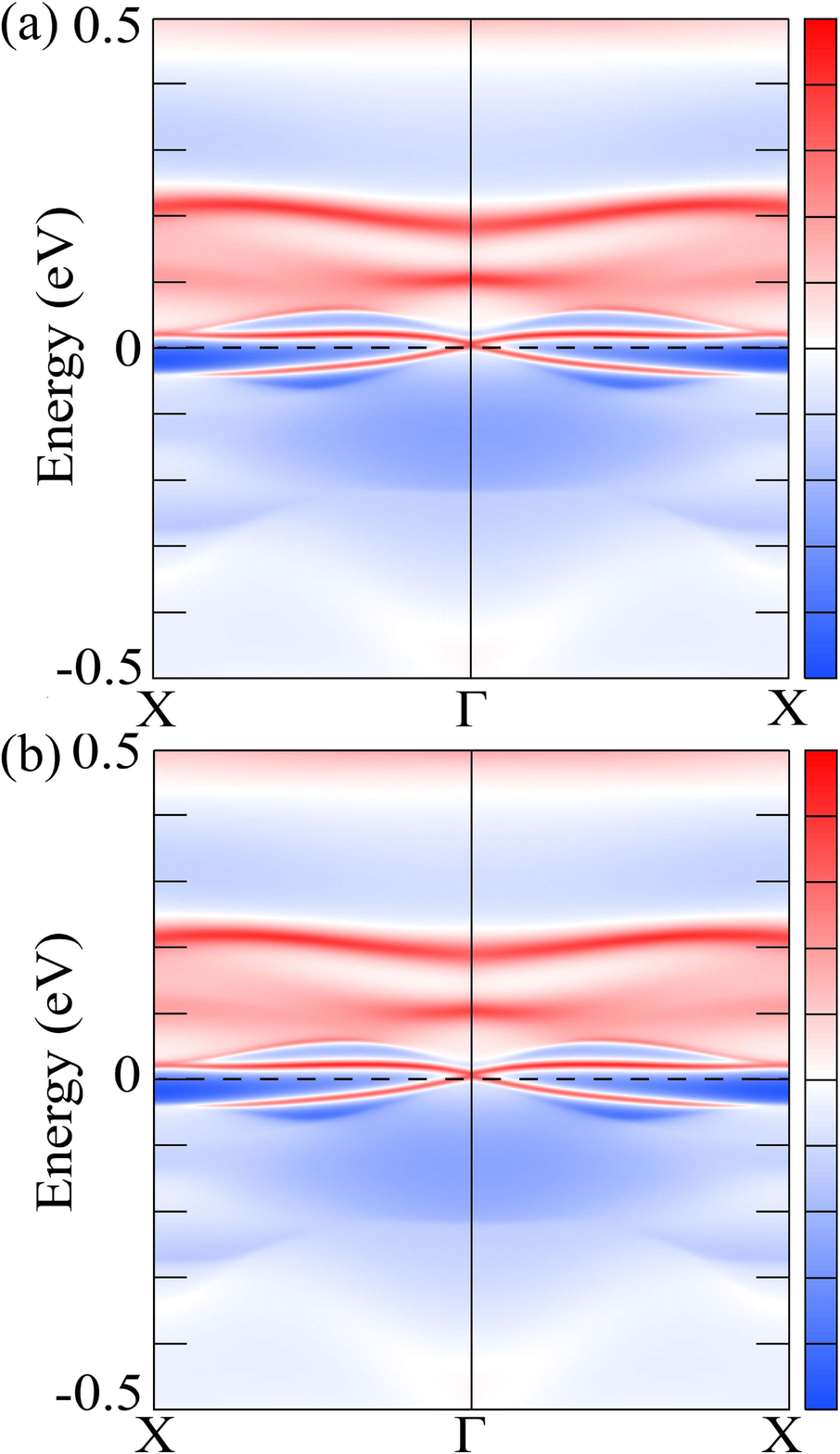}
\caption{(Color online) Edge states of (a) left and (b) right terminations in monolayer TaIrTe$_{4}$.}
\label{Fig4}
\end{figure}

According to the bulk-boundary correspondence, the nontrivial topological property of bulk accompanies with the nontrivial edge or surface states at its boundaries. The edge states of monolayer TaIrTe$_{4}$ were calculated in open boundary condition. Both the left and right terminated edges of monolayer TaIrTe$_{4}$ along $b$ axis are made up of Ta, Ir and Te atoms [Figs. \ref{Fig1}(a) and \ref{Fig1}(b)]. As a result, the edge states at the left and right terminations along $b$ axis are very similar [Figs. \ref{Fig4}(a) and \ref{Fig4}(b)] even though the two terminations are in difference with a fractional translation [Fig. \ref{Fig1}(b)]. Meanwhile, the Dirac cone located at the $\Gamma$ point is very close to the Fermi level.


\begin{figure}[!t]
\includegraphics[width=0.8\columnwidth]{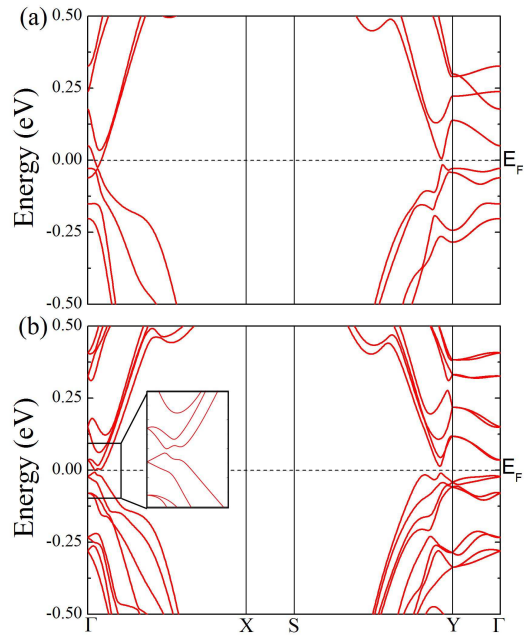}
\caption{(Color online) Band structures of bilayer TaIrTe$_{4}$ calculated (a) without and (b) with SOC. Inset shows the enlarged band structure around $\Gamma$ point.}
\label{Fig5}
\end{figure}

In comparison with monolayer TaIrTe$_{4}$, the space inversion symmetry is absent in bilayer TaIrTe$_{4}$. Bilayer TaIrTe$_{4}$ has the P1m1 symmorphic space group symmetry. The corresponding point group symmetry operations contain invariant E and mirror reflection M$_{x}$. Figure 5 shows the calculated band structure of bilayer TaIrTe$_{4}$. The bilayer appears to be a semimetal without including SOC, as shown in Fig. \ref{Fig5}(a), in which there is a band crossing along the $\Gamma$-X path as well as band inversion around the $\Gamma$ point of SBZ. With the inclusion of SOC, the spin degeneracy is lifted due to breaking space inversion symmetry. The antisymmetric SOC is so strong that it results in a large band splitting [Fig. \ref{Fig5}(b)]. Meanwhile, the band crossing along the $\Gamma$-X path opens a 8-meV gap [inset in Fig. \ref{Fig5}(b)]. Remarkably, the calculated topological invariant Z$_{2}$ is equals to 1, indicating that bilayer TaIrTe$_{4}$ is still a QSH insulator.

\begin{figure}[!t]
\includegraphics[width=0.7\columnwidth]{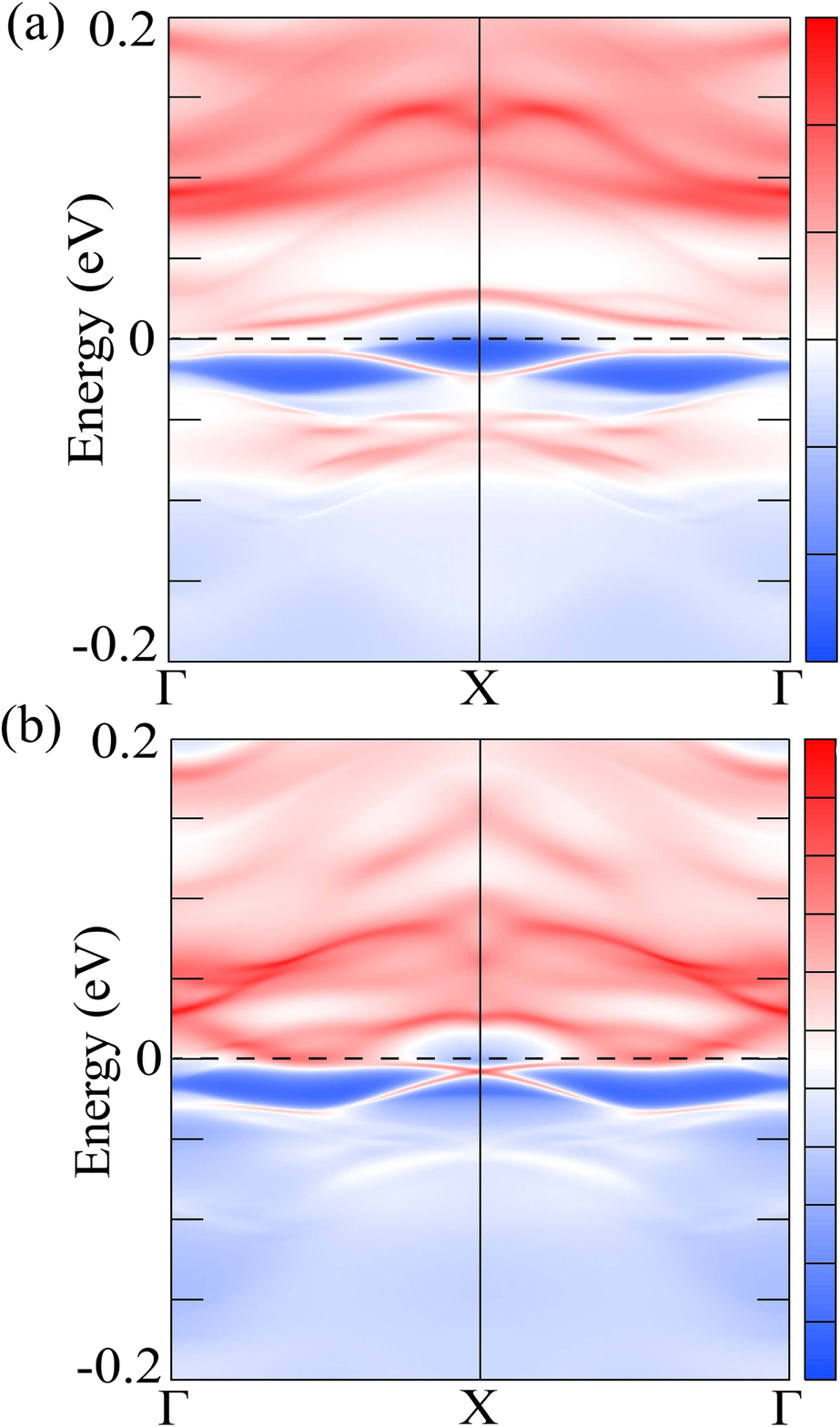}
\caption{(Color online) Edge states of (a) left and (b) right terminations in bilayer TaIrTe$_{4}$.}
\label{Fig6}
\end{figure}

We further calculated the edge states of bilayer TaIrTe$_{4}$ with open boundary condition. The calculated helical edge states of left and right terminations along $b$ axis are shown in Figs. \ref{Fig6}(a) and \ref{Fig6}(b), respectively. On this occasion, the two edge states show difference. The Dirac cone overlaps with the bulk states at the left edge, but not at the right edge. In addition,  the Dirac cone of helical edge states for bilayer TaIrTe$_{4}$ is located at the X point, instead of the $\Gamma$ point in monolayer TaIrTe$_{4}$ (Fig. \ref{Fig4}).

\begin{figure}[!t]
\includegraphics[width=0.8\columnwidth]{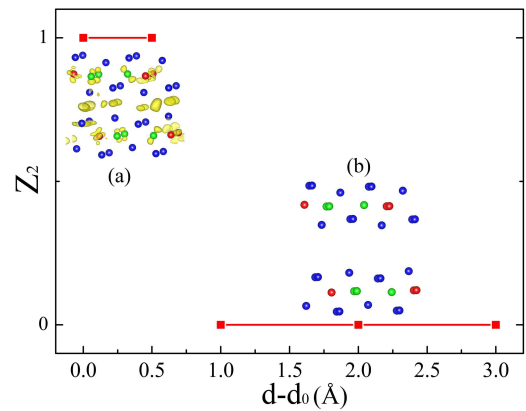}
\caption{(Color online) Topological invariant Z$_{2}$ with increasing interlayer distance for bilayer TaIrTe$_{4}$.  Differential charge densities for (a) equilibrium structure ($d=d_0$) and (b) enlarged interlayer distance ($d=d_0+2 $ \AA), respectively. Here only positive differential charge densities are illustrated.}
\label{Fig7}
\end{figure}

 Although stacking of two QSH insulators is usually considered to give rise to a trivial insulator, once there appear odd times of qualified band inversions, this may, again, make the bilayer non-trival. The calculations show that there are two bands near the Fermi level inverting once around the $\Gamma$ point [Fig. \ref{Fig5}(b)], this makes bilayer TaIrTe$_{4}$ be still a QSH insulator. The underlying reason is as follows. In many layered materials, strong interlayer wavefunction overlap of $p$ orbitals does occur, known as covalent-like quasi bonding (CLQB), which is a result of balanced dispersion attraction and, mostly, Pauli repulsion. The formation of CLQB induces broadening of valence and conduction bands and thus shows layer-dependent bandgaps and other electronic structure related properties~\cite{Qiao-nc, Zhao-am, Qiao-sb, Wang-PRB}. Previously, both elemental layers and compounds of tellurium were found to possess such covalent-like interlayer interaction~\cite{Qiao-sb, Liu-PRM}. For bilayer TaIrTe$_{4}$, the left inset of Fig. \ref{Fig7}(a) depicts a plot of interlayer differential charge densities (DCD) at the equilibrium interlayer distance $d_0$, which explicitly shows charge accumulation (covalent characteristic) at the interlayer region. As shown in Fig. \ref{Fig3}, the bandgap of monolayer TaIrTe$_{4}$ is small around the $\Gamma$ point. Thus the largely broadened valence and conduction bands of bilayer TaIrTe$_{4}$ give rise to a band inversion over the small bandgap around the $\Gamma$ point [Fig. \ref{Fig5}(b)]. In contrast, monolayer WTe$_{2}$ is a QSH insulator~\cite{Qian2014} as well as with the covalent-like interlayer bonding, but it possesses large band gaps at the time-reversal invariant points. As a result, bilayer WTe$_{2}$ is a trivial insulator, as verified by the calculations. Thus the QSH phase is a distinct property of bilayer TaIrTe$_{4}$.

The interlayer wavefunction overlap should gradually weaken as the interlayer distance $d$ increases. Therefore the bilayer will undergo a topological phase transition with the increasing interlayer distance. This analysis is confirmed by our calculation that topological invariant Z$_{2}$ of the bilayer goes from 1 to 0 in between $d-d_0=0.5$ \AA~ and 1 \AA~(Fig. \ref{Fig7}), indicating a topological nontrivial to trivial transition. While the interlayer charge sharing is clearly shown in DCD plots at distances near $d_{0}$, no appreciable charge sharing is observable at an interlayer distance $d-d_0=2$ \AA~ [Fig. \ref{Fig7}(b)], suggesting the CLQB eliminated.

$Discussion$: As analyzed above, the covalent-like interlayer interaction and the small band gap at time-reversal invariant point are key factors for the emergence of QSH phase in bilayer TaIrTe$_{4}$. Considering that the covalent-like interlayer interaction exists quite popularly in real layered materials, our study suggests that other bilayers or van der Waals (vdW) heterostructrues may also realize interesting topological phases, for example, Dirac, Weyl, and node-line semimetals, besides QSH insulator.


A very recent transport experiment found that bulk TaIrTe$_{4}$ becomes superconducting under pressure~\cite{Cai-PRB}. Similarly, the superconductivity in bulk WTe$_{2}$~\cite{Kang-nc, Pan-nc} can be induced by pressure, while the transition from 2D topological insulator to superconductor is driven in monolayer WTe$_{2}$ via the application of gate voltage~\cite{Sajadi-s}. Nontrivial topological superconductivity in monolayer and bilayer TaIrTe$_{4}$ may also be induced by gate voltage or stress. Furthermore, due to the broken inversion symmetry, Ising superconductivity~\cite{Xi-np} may be realized in bilayer TaIrTe$_{4}$.

$Summary$: Based on first-principles electronic structure calculations, we confirm that monolayer TaIrTe$_{4}$ is a QSH insulator. The Dirac cone of its robust helical edge states located at the $\Gamma$ point is very close to the Fermi level. Remarkably, due to the covalent-like interlayer interaction and the small bandgap at the time-reversal invariant $\Gamma$ point, bilayer TaIrTe$_{4}$ is still a QSH insulator. Meanwhile, a topological phase transition is observed by increasing the interlayer distance in bilayer TaIrTe$_{4}$. 

\begin{acknowledgments}

This work was supported by the National Key R\&D Program of China (Grants No. 2017YFA0302903 and No. 2018YFE0202700), the National Natural Science Foundation of China (Grants No. 11774422, No. 11774424, No. 11622437, and No. 11974422), the Strategic Priority Research Program (Chinese Academy of Sciences, CAS) (No. XDB30000000), the CAS Interdisciplinary Innovation Team, the Fundamental Research Funds for the Central Universities, and the Research Funds of Renmin University of China (Grants No. 16XNLQ01 and No. 19XNLG13). Computational resources have been provided by the Physical Laboratory of High Performance Computing at Renmin University of China.

\end{acknowledgments}


\begin{thebibliography}{}

\bibitem{Hasan} M. Z. Hasan and C. L. Kane, Rev. Mod. Phys. \textbf{82}, 3045 (2010).
\bibitem{Qi} X. L. Qi and S. C. Zhang, Rev. Mod. Phys. \textbf{83}, 1057 (2011).
\bibitem{Soluyanov-2015}A. A. Soluyanov, D. Gresch, Z. J. Wang, Q. S. Wu, M. Troyer, X. Dai, and B. A. Bernevig, Nature \textbf{527}, 495 (2015).
\bibitem{Qian2014} X. F. Qian, J. W. Liu, L. Fu, and J. Li, Science \textbf{346}, 1344 (2014).
\bibitem{Brink-PRB}K. Koepernik, D. Kasinathan, D. V. Efremov, S. Khim, S. Borisenko, B. Buchner, and J. V. D. Brink, Phys. Rev. B \textbf{93}, 201101(R) (2016).
\bibitem{Khim-PRB}S. Khim, K. Koepernik, D. V. Efremov, J. Klotz, T. Forster, J. Wosnitza, M. I. Sturza, S. Wurmehl, C. Hess, J. V. D. Brink, and B. Buchner, Phys. Rev. B \textbf{94}, 165145 (2016).
\bibitem{Hasan-nc}I. Belopolski, P. Yu, D. S. Sanchez, Y. Ishida, T. R. Chang, S. T. S. Zhang, S. Y. Xu, H. Zheng, G. Q. Chang, G. Bian, H. T. Jeng, T. Kondo, H. Lin, Z. Liu, S. Shin, and M. Z. Hasan, Nat. Commun. \textbf{8}, 942 (2017).
\bibitem{Haubold-PRB}E. Haubold, K. Koepernik, D. Efremov, S. Khim, A. Fedorov, Y. Kushnirenko, J. van den Brink, S. Wurmehl, B. B\"uchner, T. K. Kim, M. Hoesch, K. Sumida, K. Taguchi, T. Yoshikawa, A. Kimura, T. Okuda, and S. V. Borisenko, Phys. Rev. B \textbf{95}, 241108(R) (2017).
\bibitem{Zhou-PRB}X. Q. Zhou, Q. H. Liu, Q. S. Wu, T. Nummy, H. X. Li,J. Griffith, S. Parham, J. Waugh, E. Emmanouilidou,B. Shen, O. V. Yazyev, N. Ni, and D. Dessau, Phys. Rev. B \textbf{97}, 241102(R) (2018).
\bibitem{Dong-prb} X. Dong, M. Y. Wang, D. Y Yan, X. L. Peng, J. Li, W. D. Xiao, Q. S. Wang, J. F. Han, J. Ma, Y. G. Shi, and Y. G. Yao,  ACS Nano \textbf{13}, 9571 (2019).
\bibitem{Liu-PRM}C. Liu, C. S. Lian, M. H. Liao, Y. Wang, Y. Zhong, C. Ding, W. Li, C. L. Song, K. He, X. C. Ma, W. H. Duan, D. Zhang, Y. Xu, L. L. Wang, and Q. K. Xue, Phys. Rev. Mater. \textbf{2}, 094001 (2018).
\bibitem{Qiao-sb} J. S. Qiao, Y. H. Pan, F. Yang, C. Wang, Y. Chai, and W. Ji, Sci. Bull. \textbf{63}, 159 (2018).
\bibitem{Kresse-1993}G. Kresse and J. Hafner, Phys. Rev. B \textbf{47}, 558 (1993).
\bibitem{Kresse-1996} G. Kresse and J. Furthm\"uller, Comput. Mater. Sci. \textbf{6}, 15 (1996).
\bibitem{Kresse-J} G. Kresse and J. Furthm\"uller, Phys. Rev. B \textbf{54}, 11169 (1996).
\bibitem{B-PRB}P. E. Bl\"ochl, Phys. Rev. B \textbf{50}, 17953 (1994).
\bibitem{Kresse-PRB}G. Kresse and D. Joubert, Phys. Rev. B \textbf{59}, 1758 (1999).
\bibitem{Perdew-PRL}J. P. Perdew, K. Burke, and M. Ernzerhof, Phys. Rev. Lett. \textbf{77}, 3865 (1996).
\bibitem{vdw-2011} J. Klime\v s, D. R. Bowler, and A. Michaelides, Phys. Rev. B \textbf{83}, 195131 (2011).
\bibitem{Mar-1992}A. Mar, S. Jobic, and J. A. Ibers, J. Am. Chem. Soc. \textbf{114}, 8963 (1992).
\bibitem{Wu-wannier}Q. S. Wu, S. Zhang, H.-F. Song, M. Troyer, and A. A. Soluyanov, Comput. Phys. Commun. \textbf{224}, 405 (2018).
\bibitem{Zacharia-PRB}R. Zacharia, H. Ulbricht, and T. Hertel, Phys. Rev. B \textbf{69}, 155406 (2004).
\bibitem{Krukau-2006}A. V. Krukau, O. A. Vydrov, A. F. Izmaylov, and G. E. Scuseria, J. Chem. Phys. \textbf{125}, 224106 (2006).
\bibitem{Yu-PRB}R. Yu, X. L. Qi, A. Bernevig, Z. Fang, and X. Dai, Phys. Rev. B \textbf{84}, 075119 (2011).
\bibitem{Qiao-nc} J. S. Qiao, X. Kong, Z. X. Hu, F. Yang, and W. Ji, Nat. Commun. \textbf{5}, 4475 (2014).
\bibitem{Zhao-am}Y. D. Zhao, J. S. Qiao, Z. H. Yu, P. Yu, K. Xu, S. P. Lau, W. Zhou, Z. Liu, X. R. Wang, W. Ji, and Y. Chai, Adv. Mater. \textbf{29}, 1604230 (2016).
\bibitem{Wang-PRB}C. Wang, X. Y. Zhou, Y. H. Pan, J. S. Qiao, X. H. Kong, C. C. Kaun, and W. Ji, Phys. Rev. B \textbf{97}, 245409 (2018).
\bibitem{Cai-PRB}S. Cai, E. Emmanouilidou, J. Guo, X. D. Li, Y. C. Li, K. Yang, A. Li, Q. Wu, N. Ni, and L. L. Sun, Phys. Rev. B \textbf{99}, 020530(R) (2019).
\bibitem{Kang-nc}D. F. Kang, Y. Z. Zhou, W. Yi, C. L. Yang, J. Guo, Y. G. Shi, S. Zhang, Z. Wang, C. Zhang, S. Jiang, A. G. Li, K. Yang, Q. Wu, G. M. Zhang, L. L. Sun, and Z. X. Zhao, Nat. Commun. \textbf{6}, 7804 (2015).
\bibitem{Pan-nc}X. C. Pan, X. L. Chen, H. M. Liu, Y. Q. Feng, Z. X. Wei, Y. H. Zhou, Z. H. Chi, L. Pi, F. Yen, F. Q. Song, X. G. Wan, Z. R. Yang, B. G. Wang, G. H. Wang, and Y. H. Zhang, Nat. Commun. \textbf{6}, 7805 (2015).
\bibitem{Sajadi-s} E. Sajadi, T. Palomaki, Z. Y. Fei, W. J. Zhao, P. Bement, C. Olsen, S. Luescher, X. D. Xu, J. A. Folk, and D. H. Cobden, Science \textbf{362}, 922 (2018).
\bibitem{Xi-np}X. X. Xi, Z. F. Wang, W. W. Zhao, J. H. Park, K. T. Law, H. Berger, L. Forr\'o, J. Shan, and K. F. Mak, Nat. Phys. \textbf{12}, 139 (2016).
\end{thebibliography}
\end{document}